# Crystal structure and phase transitions across the metal-superconductor boundary in the SmFeAsO$_{1-x}$F$_x$ (0 ≤ *x* ≤ 0.20) family

Serena Margadonna[1*], Yasuhiro Takabayashi[2], Martin T. McDonald[2], Michela Brunelli[3], G. Wu[4], R. H. Liu[4], X. H. Chen[4] & Kosmas Prassides[2*]

[1] *School of Chemistry, University of Edinburgh, Edinburgh EH9 3JJ, UK*

[2] *Department of Chemistry, University of Durham, Durham DH1 3LE, UK*

[3] *European Synchrotron Radiation Facility, 38042 Grenoble, France*

[4] *Department of Physics, University of Science and Technology of China, Hefei, Anhui 230026, P. R. China*

*\*e-mail: serena.margadonna@ed.ac.uk, K.Prassides@durham.ac.uk*

**The fluorine-doped rare-earth iron oxyarsenides, REFeAsO$_{1-x}$F$_x$ (RE =rare earth) have recently emerged as a new family of high-temperature superconductors with transition temperatures ($T_c$) as high as 55 K (refs 1-4). Early work has provided compelling evidence that the undoped parent materials exhibit spin-density-wave (SDW) antiferromagnetic order and undergo a structural phase transition from tetragonal to orthorhombic crystal symmetry upon cooling.[5] Both the magnetic and structural instabilities are suppressed upon doping with fluoride ions before the appearance of superconductivity.[6,7] Here we use high-resolution synchrotron X-ray diffraction to study the structural properties of SmFeAsO$_{1-x}$F$_x$ (0 ≤ *x* ≤ 0.20) in which superconductivity emerges near *x* ~ 0.07 and $T_c$ increases monotonically with doping up to *x* ~ 0.20.[8] We find that orthorhombic symmetry survives through the metal-superconductor boundary well into the superconducting regime** **and the structural distortion is only suppressed at doping levels, $x \geq 0.15$ when the superconducting phase becomes metrically tetragonal. Remarkably this crystal symmetry crossover coincides with reported drastic anomalies in the resistivity and the Hall coefficient[8] and a switch of the pressure coefficient of $T_c$ from positive to negative,[9] thereby implying that the low-temperature structure plays a key role in defining the electronic properties of these superconductors.**

The possible mechanism of superconductivity in the REFeAsO$_{1-x}$F$_x$ and related REFeAsO$_{1-\delta}$ materials is currently unknown. The rapidly developing structural and electronic phenomenology points to considerable similarities with the well-established behaviour of high-$T_c$ cuprate superconductors and early theoretical work has suggested that conventional electron-phonon coupling mechanisms are not able to account for the high $T_c$, implying non-BCS origin of the pairing interactions.[10-13] The parent REFeAsO phases exhibit both a structural and a magnetic phase transition on cooling in a similar fashion to the parent cuprate phase, La$_2$CuO$_4$.[5,14] Upon doping with fluoride ions, again much like La$_{2-x}$Sr$_x$CuO$_4$, both the crystallographic and magnetic transitions are suppressed in the superconducting compositions,[6,7] while $T_c$ first increases smoothly before passing over a maximum value at an optimal level of doping. Detailed experimental mapping of the structural and electronic phase diagrams as the doping level varies is necessary before we achieve a fundamental understanding of the superconductivity mechanism.

Here we probed the temperature evolution of the structural properties of the SmFeAsO$_{1-x}$F$_x$ ($x$ = 0, 0.05, 0.10, 0.12, 0.15, and 0.20) family by high-resolution synchrotron X-ray powder diffraction and examined the diffraction profiles collected over an angular range of 1° to 40° ($d$-spacing = 22.85 to 0.59 Å) at various temperatures between 295 and 20 K. Inspection of all diffraction profiles at room temperature readily reveals the tetragonal (T) unit cell (space group $P4/nmm$) established before for other REFeAsO systems.[5,14] Therefore all the SmFeAsO$_{1-x}$F$_x$ compositions studied here are isostructural and adopt the layered ZrCuSiAs-type structure, featuring alternating tetrahedrally coordinated Sm-O/F and Fe-As layers along the crystallographic $c$ axis. Rietveld analysis of the room temperature diffraction profiles proceeded smoothly for all compositions, revealing a monotonic decrease in both lattice constants with increasing doping level, $x$ (SmFeAsO: $a_T$ = 3.93880(2) Å, $c_T$ = 8.51111(7) Å; SmFeAsO$_{0.80}$F$_{0.20}$: $a_T$ = 3.93254(4) Å, $c_T$ = 8.4842(1) Å). The response of the lattice metrics to F substitution is strongly anisotropic with the interlayer spacing showing a significantly larger contraction than the intralayer dimensions with increasing $x$ ($\partial \ln c_T/\partial x \sim 1.6 \times 10^{-2}$, $\partial \ln a_T/\partial x \sim 0.8 \times 10^{-2}$).

However, the structural behaviour of the SmFeAsO$_{1-x}$F$_x$ compositions is very different on cooling. No reflections violating tetragonal extinction rules are evident for the heavily-doped compositions with $x$ = 0.15 and 0.20 (Fig. 1e and 1f), in which both lattice constants, $a$ and $c$ decrease smoothly with their crystal structure remaining strictly tetragonal down to 20 K (Fig. 2e and 2f). The rate of contraction, d$\ln a$/d$T$ and d$\ln c$/d$T$ at ~5 and ~18 ppm K$^{-1}$ for the $a$ and $c$ lattice constants, respectively is considerably anisotropic and leads to a gradual decrease of the ($c/a$) ratio with decreasing temperature. This behaviour is in sharp contrast to the observed thermal structural response of the SmFeAsO$_{1-x}$F$_x$ ($x$ = 0, 0.05, 0.10, and 0.12) compositions. In these systems, the tetragonal structure is initially robust upon cooling showing a normal contraction of the lattice parameters and interatomic distances. However, as the samples are cooled further, all $hkl$ ($h$, $k \neq 0$) reflections in the diffraction profiles begin first to broaden before splitting at a characteristic temperature, $T_s$ (Fig. 1a-1d) thereby providing the signature of the onset of a structural transformation of the high-temperature tetragonal structure.[5,14] Rietveld refinements of the low-temperature diffraction profiles confirm the adoption of the same orthorhombic (O) superstructure of lattice dimensions, $b_O > a_O \sim a_T\sqrt{2}$ and $c_O \sim c_T$ (space group *Cmma*) for all $0 \leq x \leq 0.12$ compositions (Fig. 2a-2d). No discontinuity is observed at $T_s$ in the thermal response of either the lattice constant, $c$ or the normalised unit cell volume, $V$. Notably as the doping level, $x$ increases, both the transition temperature, $T_s$ (130 K for $x = 0$ to ~50 K for $x = 0.10$ and 0.12) and the magnitude of the orthorhombic strain coefficient, $s = (b_O - a_O)/(b_O + a_O)$ (~$2.5 \times 10^{-3}$ at 20 K for $x = 0$ to ~$1.1 \times 10^{-3}$-$1.3 \times 10^{-3}$ at 20 K for $x = 0.10$ and 0.12) decrease smoothly. The results of the final Rietveld refinements at 20 K for the SmFeAsO$_{1-x}$F$_x$ compositions with $x = 0$, 0.10, and 0.20 are shown in Supplementary Fig. S1a-S1c with the fitted parameters summarised in Supplementary Table S1.

The most prominent point arising from the results of the present structural refinements as a function of both temperature and composition is the survival of the orthorhombic crystal symmetry in SmFeAsO$_{1-x}$F$_x$ well beyond the onset of superconductivity. Crossing the metal-to-superconductor boundary at $x \sim 0.07$ is not accompanied by the complete suppression of the orthorhombic-to-tetragonal structural phase transition and, as for both $x = 0.10$ and 0.12 compositions studied here $T_s > T_c$, both superconducting phases are orthorhombically distorted (Fig. 3). Although with increasing $x$, $T_s$ is shifting continuously to lower temperature values, the tetragonal symmetry in the superconducting state does not appear until well into the $T_c$ versus $x$ superconducting dome at $x = 0.15$.

At first sight, given that $T_c$ in the SmFeAsO$_{1-x}$F$_x$ family increases smoothly between $x \sim 0.07$ and 0.20, it may appear that the orthorhombic-to-tetragonal crossover is not reflected in the electronic properties despite the clear signature of the structural transformation in the temperature dependence of the resistivity ($T_s$ coincides with the temperature at which the first derivative of the temperature-dependent resistivity, d$\rho$/d$T$ shows a maximum, Fig. 3) and in the renormalisation of the bonding interactions within the conducting Fe-As slabs that accompany it (*vide infra*). However, here we recall two additional significant experimental observations already established for the SmFeAsO$_{1-x}$F$_x$ family that point towards the existence of a criticality hidden under the smoothly shaped superconducting dome at a doping level, $x \sim 0.14$: (i) the temperature dependence of the resistivity is linear at high temperatures (low temperatures just above $T_c$) for $x < 0.14$ ($x > 0.14$); this differing temperature evolution is accompanied by a drop in carrier density as observed by the pronounced rise in the Hall coefficient,[8] and (ii) the superconducting response to pressure is drastically different for compositions straddling the $x \sim 0.14$ doping level (Fig. 3); while for $x < 0.14$ the pressure coefficient, $\partial \ln T_c/\partial P$ of SmFeAsO$_{1-x}$F$_x$ is strongly positive, it switches sharply to negative at $x > 0.14$.[9] The observation that these pronounced anomalies in the electronic properties coincide exactly with the crossover from orthorhombic ($x < 0.14$) to tetragonal ($x > 0.14$) symmetry for the superconducting phase points towards a key role played by the structural order in determining the bonding interactions within the conducting Fe-As slabs and the electronic properties of the SmFeAsO$_{1-x}$F$_x$ superconductors.

Fig. 4c shows the doping dependence at 20 K of selected crystallographic bond distances and angles. Gradual substitution of oxide by fluoride ions in the charge-reservoir Sm-O slab is accompanied by a gradual increase in the Sm-O/F distances,

Focusing on the conducting Fe-As layer, we find that the thickness of the As-Fe-As slab (Fig. 4a) shows a clear discontinuity in the vicinity of the orthorhombic-to-tetragonal structural crossover at $x \sim 0.12$. This anomalous response is even more clearly evident in the $x$ dependence of the Fe-As-Fe angles (Fig. 4b). These initially show a gradual increase in the orthorhombic phase as the doping level $x$ increases. However, the suppression of the structural transition and the stabilisation of the tetragonal phase at $x = 0.15$ is accompanied by a well-defined reduction in the magnitude of the Fe-As-Fe angles. As the geometry of the $AsFe_4$ units (Fig. 4b) sensitively controls both the Fe near- and next-near-neighbour exchange interactions[15] and the width of the electronic conduction band,[7,16] the structural discontinuities near the critical composition, $x \sim 0.14$ should be related with the observed electronic anomalies well within the superconducting dome.

Powder neutron diffraction studies on the $CeFeAsO_{1-x}F_x$ family[7] have provided evidence that the magnetic SDW long range order in the parent material is rapidly suppressed upon doping and disappears at a doping level, $x \sim 0.06$ just before superconductivity emerges. Given the criticality in the structural, electronic and conducting properties at $x \sim 0.14$ revealed for the $SmFeAsO_{1-x}F_x$ family here, it will be intriguing to search for effects of magnetic origin and establish the magnetic response of the normal state in the fluorine-doped rare-earth iron oxyarsenide families well beyond the compositional onset for superconductivity.

**METHODS**

**Sample preparation**

Polycrystalline samples with nominal composition SmFeAsO$_{1-x}$F$_x$ ($0 \leq x \leq 0.20$) were synthesised by conventional solid state reactions using high-purity SmAs, SmF$_3$, Fe, and Fe$_2$O$_3$, as described elsewhere.[8] The samples were characterised by powder X-ray diffraction and temperature-dependent resistivity and *dc* magnetisation measurements.

**Synchrotron X-ray diffraction**

For the synchrotron X-ray diffraction measurements, the SmFeAsO$_{1-x}$F$_x$ ($x$ = 0, 0.05, 0.10, 0.12, 0.15, and 0.20) samples were sealed in thin-walled glass capillaries 0.5 mm in diameter. With each sample inside a continuous-flow cryostat, high-statistics synchrotron X-ray powder diffraction data ($\lambda$ = 0.399861 Å, $2\theta$ = 1° to 50°) were collected at 20 and 200 K in continuous scanning mode with the high-resolution multianalyser powder diffractometer on beamline ID31 at the European Synchrotron Radiation Facility (ESRF), Grenoble, France. Lower statistics diffraction profiles were also recorded on cooling at numerous temperatures between 295 and 20 K over a shorter angular range ($2\theta$ = 1° to 40°). Data analysis was performed with the GSAS suite of Rietveld analysis programmes.

**Supplementary Information** is linked to the online version of the paper at www.nature.com/nature.

**Acknowledgements** We thank the ESRF for access to the synchrotron X-ray facilities.

**Competing interests statement** The authors declare that they have no competing financial interests.

**Correspondence** and requests for materials should be addressed to S.M. or K.P.

**Figure 1 | Structural characterisation of the SmFeAsO$_{1-x}$F$_x$ family.** Selected region of the high-resolution synchrotron X-ray powder diffraction profiles of SmFeAsO$_{1-x}$F$_x$ showing the temperature evolution of the (220)$_T$ Bragg reflection ($\lambda$ = 0.39986 Å). **a**, $x = 0$; **b**, $x = 0.05$; **c**, $x = 0.10$; **d**, $x = 0.12$; **e**, $x = 0.15$; and **f**, $x = 0.20$. On cooling, the tetragonal peak splits into a doublet [(040)$_O$, (400)$_O$] for $x = 0 - 0.12$, while no detectable splitting is found for $x = 0.15$ and 0.20 even at the ultrahigh resolution of the present data. The width of the (220)$_T$ reflection for $x = 0, 0.5$ begins to increase before the onset of the tetragonal-to-orthorhombic structural transition (at 150 and 130 K, respectively) providing the signature of precursor strain effects associated with the development of local structural inhomogeneities.

**Figure 2 | Temperature evolution of the structural parameters of the SmFeAsO$_{1-x}$F$_x$ family. a**, $x = 0$; **b**, $x = 0.05$; **c**, $x = 0.10$; **d**, $x = 0.12$; **e**, $x = 0.15$; and **f**, $x = 0.20$. The lattice constants are derived from Rietveld refinements of high-resolution synchrotron X-ray powder diffraction data. Green circles label the interlayer $c$ lattice constant (right scale). Red circles label the in-plane $a$ and $b$ lattice constants (left scale). In **a-d**, the $a$ and $b$ lattice constants are divided by $\sqrt{2}$ at temperatures below the tetragonal-to-orthorhombic transition.

**Figure 3 | Structural and electronic phase diagram of the SmFeAsO$_{1-x}$F$_x$ family.** The red squares mark the superconducting transition temperatures, $T_c$, the green circles the tetragonal-to-orthorhombic structural transition, $T_s$, and the blue triangles the temperature at which the first derivative of the resistivity[8] with respect to temperature, d$\rho$/d$T$ displays a maximum, $T_{max,d\rho/dT}$ (bottom panel). The top panel shows the doping level dependence of the pressure coefficient of $T_c$, dln$T_c$/d$P$.[9] The shaded bars near $x \sim 0.14$ mark the boundary for different behaviour of the temperature-dependent resistivity.[8]

**Figure 4 | Structural parameters of the SmFeAsO$_{1-x}$F$_x$ family as a function of doping. a,** Schematic diagram of the crystal structure of SmFeAsO$_{1-x}$F$_x$. **b,** Geometry of the AsFe$_4$ units and definition of the three (two) distinct Fe-As-Fe bond angles for the orthorhombic (tetragonal) crystal structure. **c,** Doping dependence of selected bond distances and angles at 20 K.

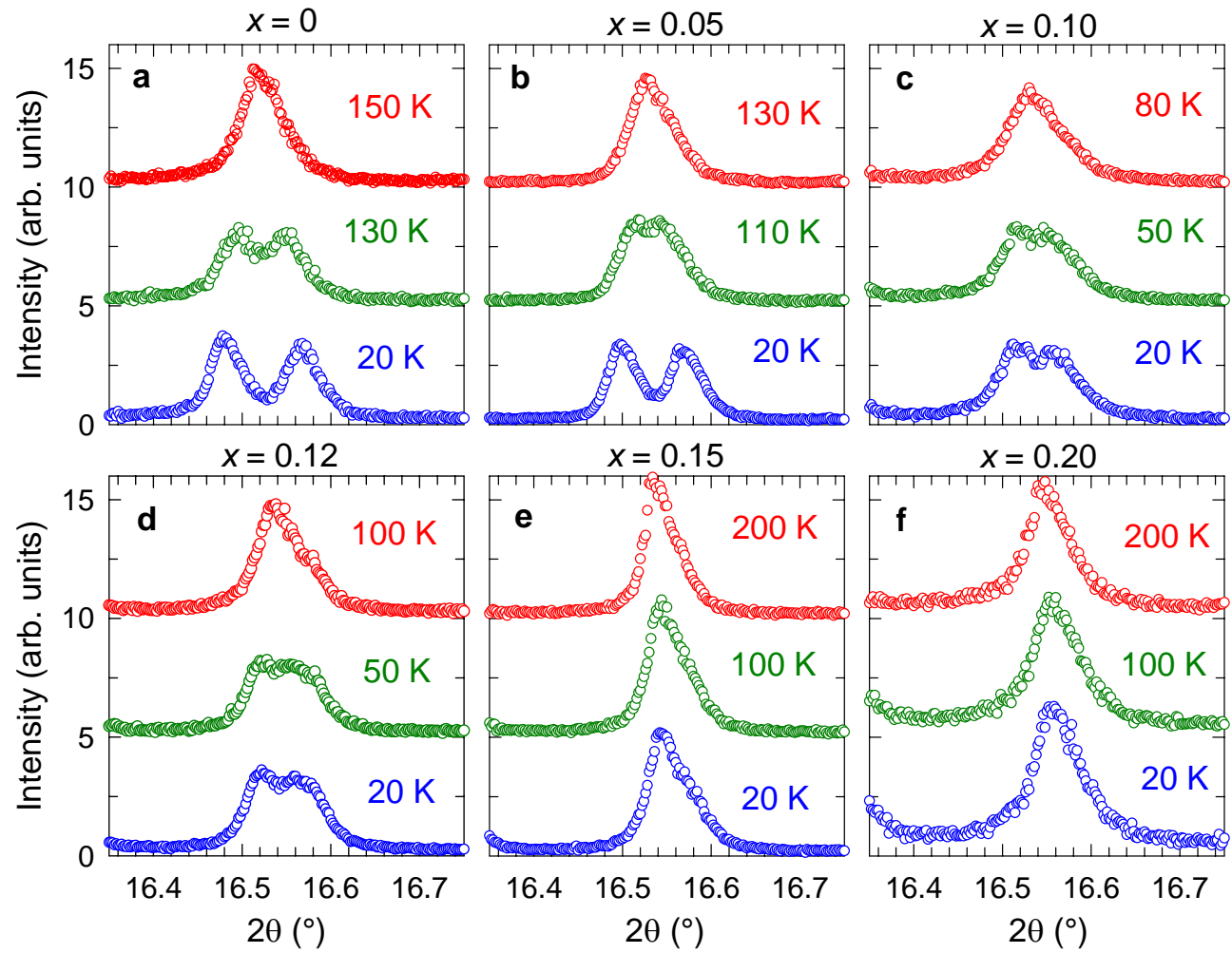

**Figure 1**

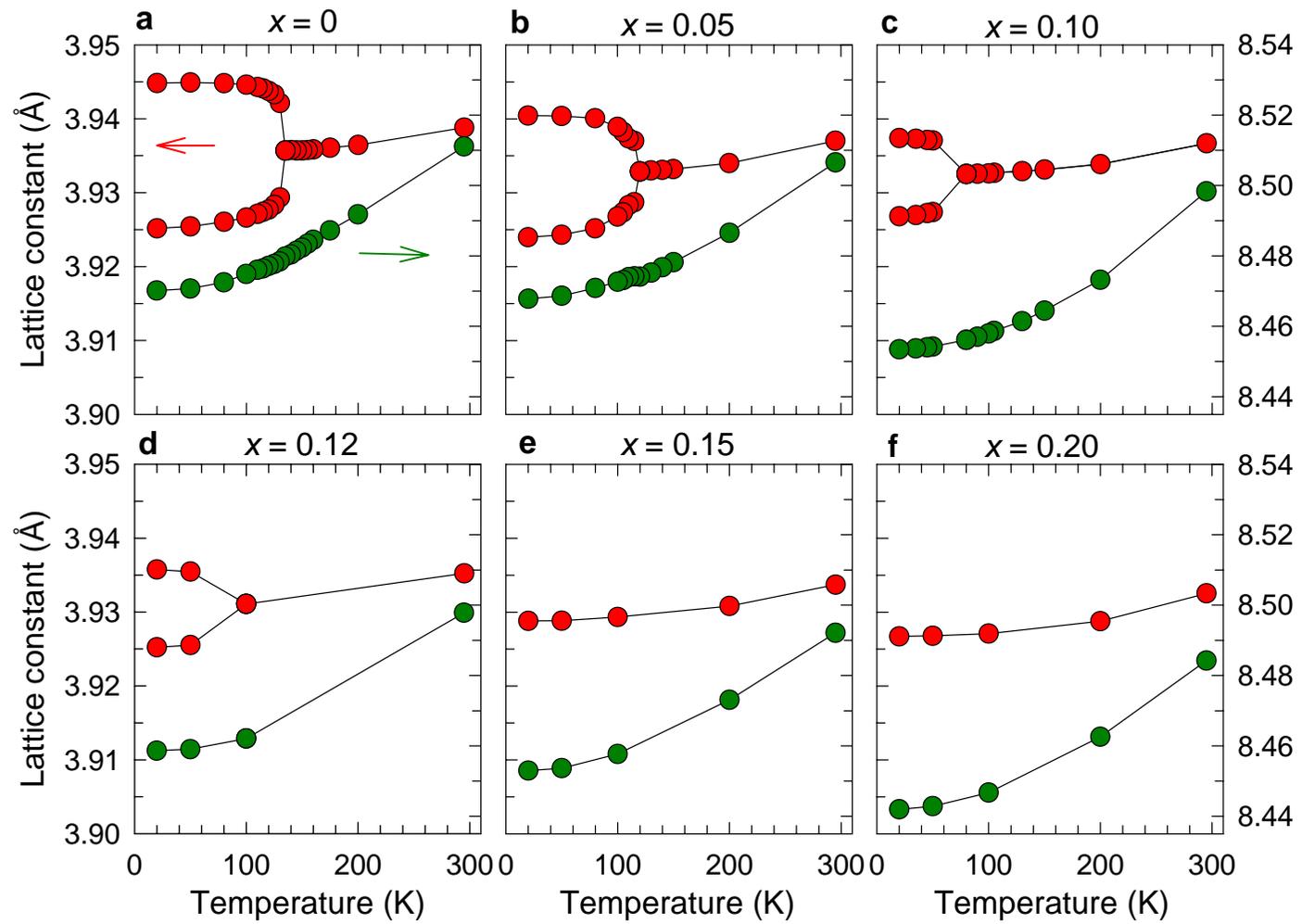

**Figure 2**

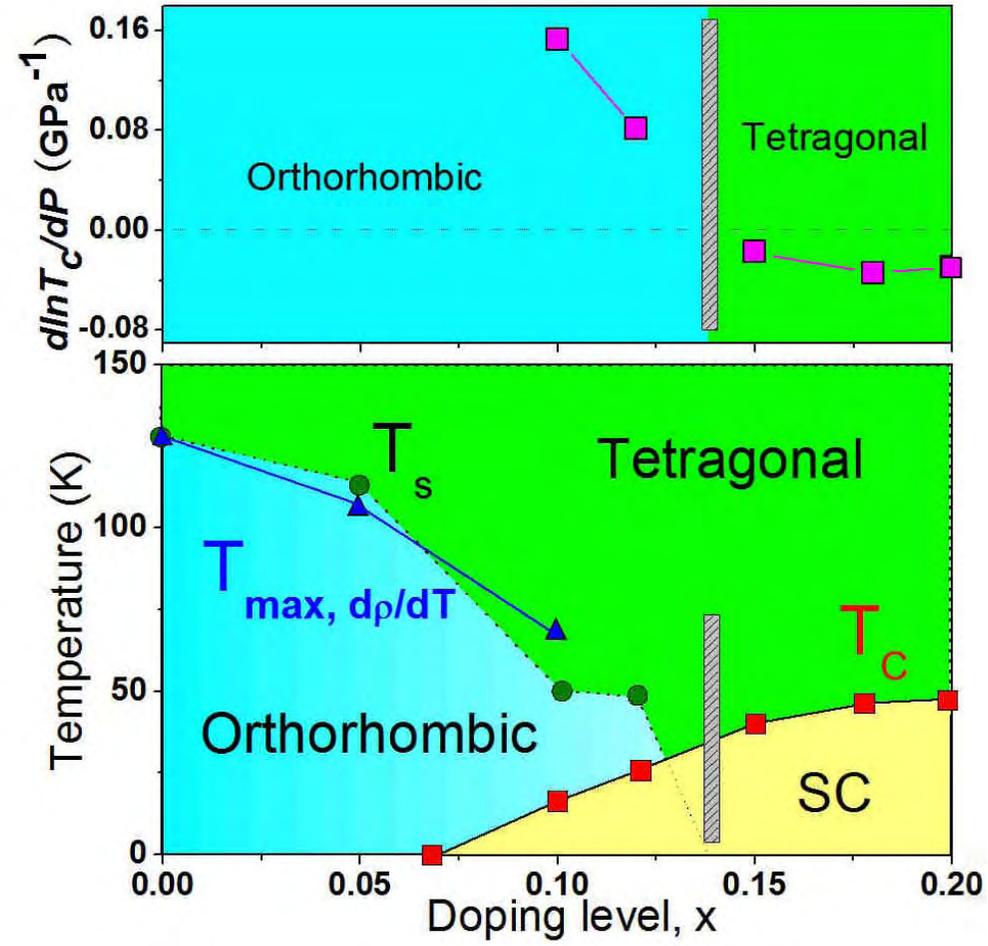

**Figure 3**

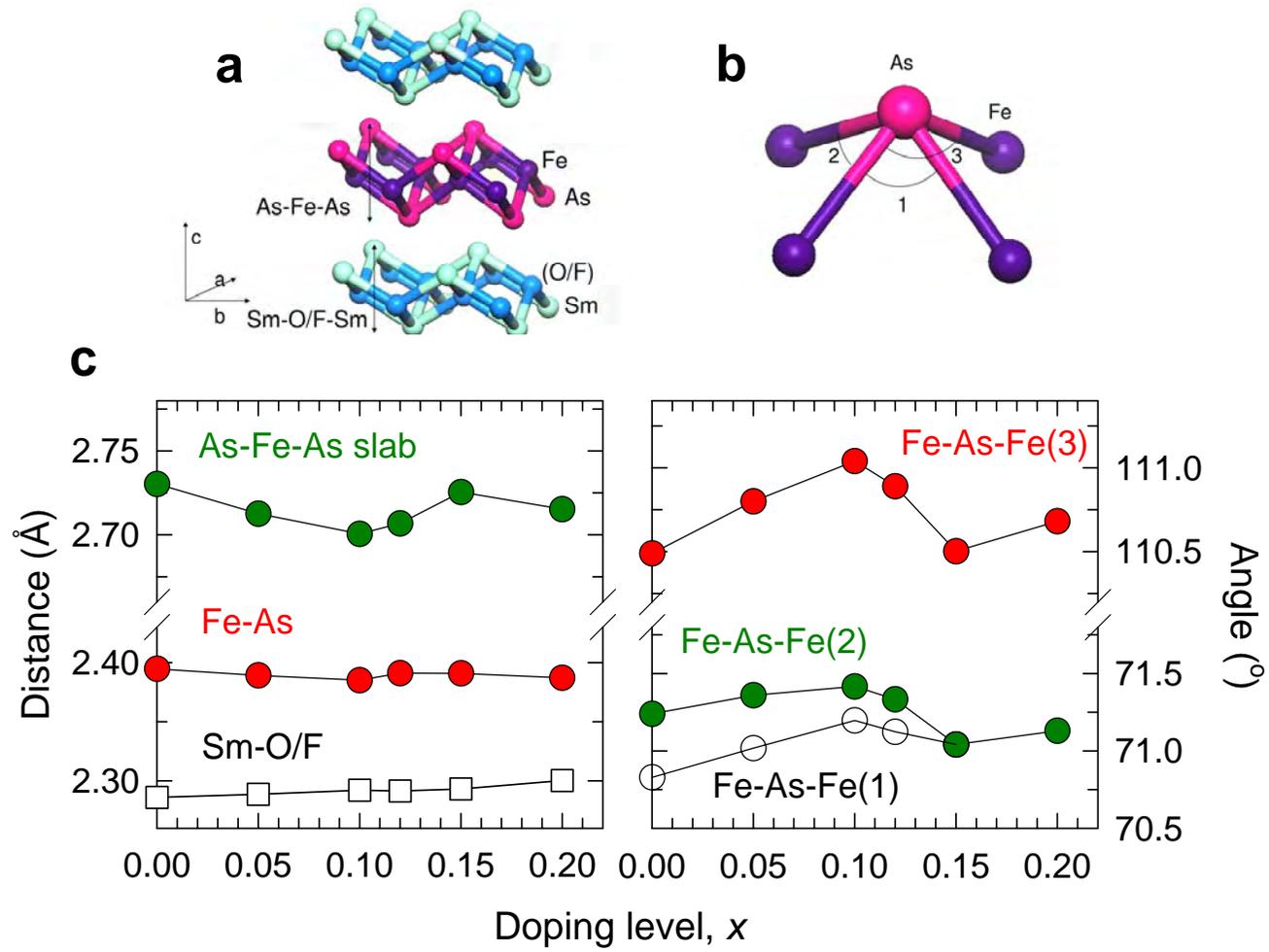

**Figure 4**